# The boron-hydrogen-phosphorus tri-elements co-doped stable N-type single crystalline Diamond

*Hongjia Bi[1&], Shaoqi Huang[1&], Feiteng Wu[1], Jiarui Guo[1], Minhui Yang[1], Yunzhen Wu[1], Mengze Zhao[3], Kaihui Liu[3] and Shisheng Lin[1,2]\**

Shisheng Lin

College of Information Science and Electronic Engineering, Zhejiang University

State Key Laboratory of Modern Optical Instrumentation, Zhejiang University

Zhejiang University, Hangzhou, 310027, China

Email: shishenglin@zju.edu.cn.

Hongjia Bi, Shaoqi Huang, Feiteng Wu, Jiarui Guo, Minhui Yang, Yunzhen Wu

College of Information Science and Electronic Engineering, Zhejiang University

Zhejiang University, Hangzhou, 310027, China

& These authors contributed equally to this work and should be considered co-first authors.

Mengze Zhao, Kaihui Liu

State Key Laboratory for Mesoscopic Physics, Frontiers Science Center for Nano-optoelectronics, School of Physics, Peking University

Peking University, Beijing, 100871, China

**Abstract**

Diamond is an outstanding semiconductor for extreme electronics, yet reproducible n-type doping remains a long-standing challenge. Here we demonstrate stable n-type single-crystal diamond grown in a single step by a precisely controlled boron-hydrogen-phosphorus co-doping strategy. Hall measurements yield electron concentrations up to $1.0 \times 10^{19}$ cm$^{-3}$ with a resistivity as low as 0.249 $\Omega \cdot$cm. Secondary-ion mass spectrometry shows that tri-elements doping is the key for achieving n-type conductivity as the electron density exceeds the incorporated phosphorus concentration and is the same level of that of hydrogen and boron concentrations, supporting a donor mechanism beyond an isolated substitutional phosphorus or just boron-hydrogen co-doping. Temperature-dependent photoluminescence (PL) reveals this tri-elements codoping method induces the impurity band, and the donor level is quite shallow around 61.6 meV, consistent with the temperature dependent resistance measurements. Moreover, the co-doped diamond also exhibits strong ultraviolet emission near 270-285 nm, and the internal quantum efficiency is estimated to be 69.4%, while the undoped diamond or only boron doped diamond shows negligible UV emission. These results establish a practical route to low-resistance high luminous n-type diamond and its based chips.

# 1. Introduction

Human culture highly put emphasis on the diamond material and recently diamond is reconsidered as the platform of next electronic chip. Diamond is a carbon semiconductor with a wide bandgap larger than 5.0 eV and extremely high thermal conductivity, which enables diamond chips to operate at higher temperatures and voltages[1]. Diamond can be doped using methods such as ion implantation, high-temperature and high-pressure processing, and microwave plasma chemical vapor deposition (MPCVD)[2]. MPCVD can grow high-quality, large-size single-crystalline diamond and dope impurities into the diamond lattice during growth, making it compatible with modern integrated circuit technology. In contrast to the achievement in fabricating p-type diamond by a boron substitution shallow acceptor of 0.37 eV, n-type single crystal diamond is rarely reported, as phosphorus is a deep donor with energy level of 0.57 eV below the conduction band[3]. On the other hand, n-type phosphorus-doped diamond faces the challenge of low phosphorus doping efficiency due to the donor level[3,4]. The S-P co-doping method has been investigated for achieving n-type diamond, although with deep donor level larger than 500 meV[5,6]. Diamond is also recognized as the first semiconductor that can become superconductive[7]. Recently, we have achieved superconductive boron-nitrogen co-doped diamond[8]. We noted that B-N co-doping can also be applied for achieving n-type diamond, although with a low carrier concentration[9,10]. On the other hand, some calculations suggest that a B-P pair can create a shallower donor-like level than isolated phosphorus, due to local lattice relaxation and Coulomb stabilization[11]. However, most studies indicate the B-P complex tends to be neutral or deep, not a truly shallow donor (activation energies still on the order of hundreds of meV). Moreover, some calculations show that $BH_2$ and phosphorus-H complexes can act as shallow donors in the diamond lattice[5,6]. In contrast, it is also frequently claimed that $BH_2$ and phosphorus-H complexes should be deep donors[12]. Although the n-type doping of diamond is indispensable for achieving complementary metal-oxide-semiconductor (CMOS) based chips, achieving the n-type diamond is a very complicated and confusing puzzle, which must be addressed through a clear logical approach.

Historically, similar to the difficulties of realizing shallow donors in diamond, p-type GaN and ZnO have been extremely difficult to achieve, as they also suffer from the doping asymmetry problem[13–17]. Theoretical studies suggest that one can first introduce mutually passivated donor-acceptor complexes to induce an impurity band, and then apply effective doping to this impurity band, thereby markedly reducing the impurity ionization energy[13–17]. Taking diamond as an example, passive (B-H) complexes can create an unoccupied impurity band below the conduction-band minimum; further introducing excess H preferentially "dopes" this impurity band and generates shallower donor levels. This theoretical framework also explicitly predicts that once the defect/impurity concentration required for impurity-band formation exceeds the percolation threshold and enters the heavy-doping regime, the broadening of the impurity band will further reduce the ionization energy, potentially accompanied by a certain degree of bandgap narrowing (which can be compensated via compositional tuning)[17]. Actually, hydrogen is an important element both in GaN and ZnO for formation of acceptors[18]. However, valid experiments that introduce shallow $BH_2$-related donors into the diamond lattice in a single step have been lacking. Although there are ongoing debates on whether boron-hydrogen complexes can serve as shallow donors, the in-situ experimental realization of n-type diamond via boron-hydrogen doping has not been demonstrated[12,19,20]. It has been recognized that sample inhomogeneity can lead to erroneous Hall-effect measurements[21,22]. Following this method, we introduce the phosphorus into the diamond lattice, which facilitates the formation of boron-hydrogen shallow donors. Herein, we have achieved n-type diamond with carrier concentration over $1\times10^{19}$ cm$^{-3}$ by phosphorus-boron-hydrogen co-doping method. We also find that the co-doping method can narrow the band gap of diamond up to 4.6 eV-4.7 eV, which is consistent with the theoretical prediction and promotes the formation of shallow donors in diamond[17]. The study firmly supports the shallow donor nature of BH$_2$ and the phosphorus doping brings more possibilities for the formation of BH$_2$ complexes. This research will open a new avenue for the realization of diamond-based PN junctions and high-power electronic devices.

## 2. Results and Discussions

Table 1 shows the carrier concentration of n-type diamond, where carrier concentration over $5 \times 10^{18}$ cm$^{-3}$ has been repeatedly found. However, for the n-type diamond with carrier concentration over $1 \times 10^{19}$ cm$^{-3}$, the mobility is usually lower than 1.0 cm$^2 \cdot$V$^{-1} \cdot$s$^{-1}$, which indicates that the electrons are strongly scattered. Assuming the electrons are scattered by the local hole produced by boron substituted carbon, the Hall coefficient $R_h$ can be described as:

$$R_{\mathrm{h}} = \frac{V_{\mathrm{H}} \times t}{I \times B} = \frac{p u_{\mathrm{p}}{}^2 - n u_{\mathrm{n}}{}^2}{\mathrm{e}(p u_{\mathrm{p}} + n u_{\mathrm{n}})}$$

Where $V_{\mathrm{H}}$ represents the Hall voltage, $t$ is the sample thickness, $I$ is the applied current, $B$ is the magnetic field, and $n$ and $u_{\mathrm{n}}$ represents the electron concentration and mobility, respectively, and the $p$ and $u_{\mathrm{p}}$ represents the hole concentration and mobility, respectively. The $t$ is fixed as 10 µm, $B$ is fixed as 0.5 T. Fortunately, the $u_{\mathrm{n}}$ and $u_{\mathrm{p}}$ for intrinsic diamond are on the order of 4500 cm$^2 \cdot$V$^{-1} \cdot$s$^{-1}$ for electrons and 2000 cm$^2 \cdot$V$^{-1} \cdot$s$^{-1}$ for holes[1]. Thus, for the Hall-effect measurements, if $n \gg p$, the Hall effect is still valid for achieving negative sign, which indicates the n-type conductivity[13]. Thus, in this case, we should emphasize that n-type diamond has been successfully obtained. In contrast, we noticed that there are some reports of n-type diamond with carrier concentration over $10^{20}$ cm$^{-3}$ and even $10^{21}$ cm$^{-3}$[23], which has a very low mobility even below 0.1 cm$^2$ V$^{-1}$ s$^{-1}$[24]. We should emphasize that the competition between the holes and electrons at the grain boundary and electrons can lead to an extremely high electron concentration and extremely low mobility, which is not the intrinsic behavior of the n-type diamond[13]. In our case, the single crystalline nature of diamond ensures the validity of accurate carrier concentration measurements. The stability of the n-type diamond is firmly demonstrated by repeated Hall-effect measurements, shown in the Table S1.

**Table 1. Hall measurement results for n-type and p-type diamond samples.**

| Sample | Hall coefficient (cm³/C) | Resistivity (Ω·cm) | Carrier concentration (cm⁻³) | Mobility (cm²·V⁻¹·s⁻¹) |
|--------|--------------------------|--------------------|------------------------------|------------------------|
| D241125 | -1.17 | 0.502 | $-5.33 \times 10^{18}$ | 2.29 |
| D241124 | -2.47 | 0.911 | $-2.53 \times 10^{18}$ | 2.70 |
| D250924 | +1.75 | 0.870 | $+3.57 \times 10^{18}$ | 2.01 |
| D251009 | +0.869 | 0.318 | $+7.19 \times 10^{18}$ | 2.73 |
| D250114 | -0.613 | 0.249 | $-1.02 \times 10^{19}$ | 2.45 |
| D251010 | -0.875 | 1.33 | $-7.14 \times 10^{18}$ | 0.660 |
| D251015 | -2.72 | 0.755 | $-2.30 \times 10^{18}$ | 3.60 |

Diamond possesses multiple vibronic absorption peaks, which induce many phonon modes. Figure. 1a shows the XRD spectrum of n-type diamond. In the inset of Figure. 1a, the XRD peak at 2θ ≈ 120° splits into two peaks at 119.51° and 119.94° due to excitation by Cu K$_\alpha$ (0.154 nm) and Cu K$_\beta$ (0.139 nm) radiation, respectively. The standard (400) lattice spacing is 0.0892 nm, corresponding to an XRD peak position of 118.1° under Cu K$_\alpha$ excitation, which is smaller than the 119.51° measured for n-type diamond. This indicates that the lattice expansion has been caused by the phosphorus doping, which is consistent with the Raman measurements. The tensile strain is calculated for benefiting the donor incorporation of phosphorus, which can reduce the donor ionization energy[25]. Figure. 1b shows the Raman spectrum of n-type diamond samples, p-type diamond samples and diamond substrate . In Figure. 1b, all diamond samples exhibited a sharp Raman peak near 1332 cm⁻¹ ($\omega_D$), which originates from the in-plane vibration of sp³-hybridized carbon atoms. Meanwhile the n-type diamond sample exhibited a broad Raman peak at 1410 cm⁻¹ ($\omega_N$), which may originate from BH$_2$ related defective complexes.[9] On the other hand, the tensile strain may be responsible for the 1410 cm$^{-1}$ Raman peak, as the doublet band of diamond anvil has been found after ultrahigh-pressure treatment[26]. Meanwhile, we also observed that the n-type diamond sample exhibited a higher Raman peak intensity at 1410 cm$^{-1}$ compared to the p-type diamond samples as shown in Figure. 1b. Let $I_{\omega_D}$ represent the area of the Raman peak at $\omega_D$, $I_{\omega_N}$ represent the area of the Raman peak at $\omega_N$, and $I_{\omega_N}/I_{\omega_D}$ represents the intensity ratio of the two Raman peaks. The $I_{\omega_N}/I_{\omega_D}$ values for the diamond samples are shown in Table 2. The $I_{\omega_N}/I_{\omega_D}$ values for n-type diamonds are all greater than 1, while those for p-type diamonds are all less than 1, which may be related to the content of $BH_2$ complexes. Figure. 1c shows the XPS results of the n-type diamond sample D241124. In Figure. 1c, the boron 1s peak of the n-type diamond sample D241124 is observed at 189.62 eV, and the phosphorus 2p peak is observed at 132.01 eV. The peak of phosphorus 2p should include the P 2p1/2 at 133 eV and P 2p3/2 at a binding energy of 135 eV[27]. The presence of characteristic B 1s and P 2p peaks indicates the effective doping of boron and phosphorus. The comparison between the XPS peaks for the boron and phosphorus is shown in Figure. S1, where the slightly difference could reveal the detailed difference of atomic conFigureurations between them. Figure. 1d shows the ultraviolet absorption spectra of the n-type diamond samples D241124, as well as the p-type diamond sample D251009. It can be clearly observed that all diamond samples exhibit pronounced absorption in the 230 - 80 nm wavelength range, with an absorption peak centered at approximately 260 nm. This behavior indicates that, in n-type diamond films, impurities introduce localized energy states within the band gap, enabling partial absorption of photons with energies lower than the intrinsic band gap of diamond. The corresponding energy band gap between impurity level and valence band is estimated to be 4.77 eV, which is in accordance with impurity band formation predicted theoretically[17]. The absorption spectrum of more samples for both n-type and p-type diamond aslo shows the impurity band related continuous absorption band in the 230-280 nm wavelength range as shown in Figure. S2, which demonstrates the impurity band has an intrinsic width as a continuous energy band.

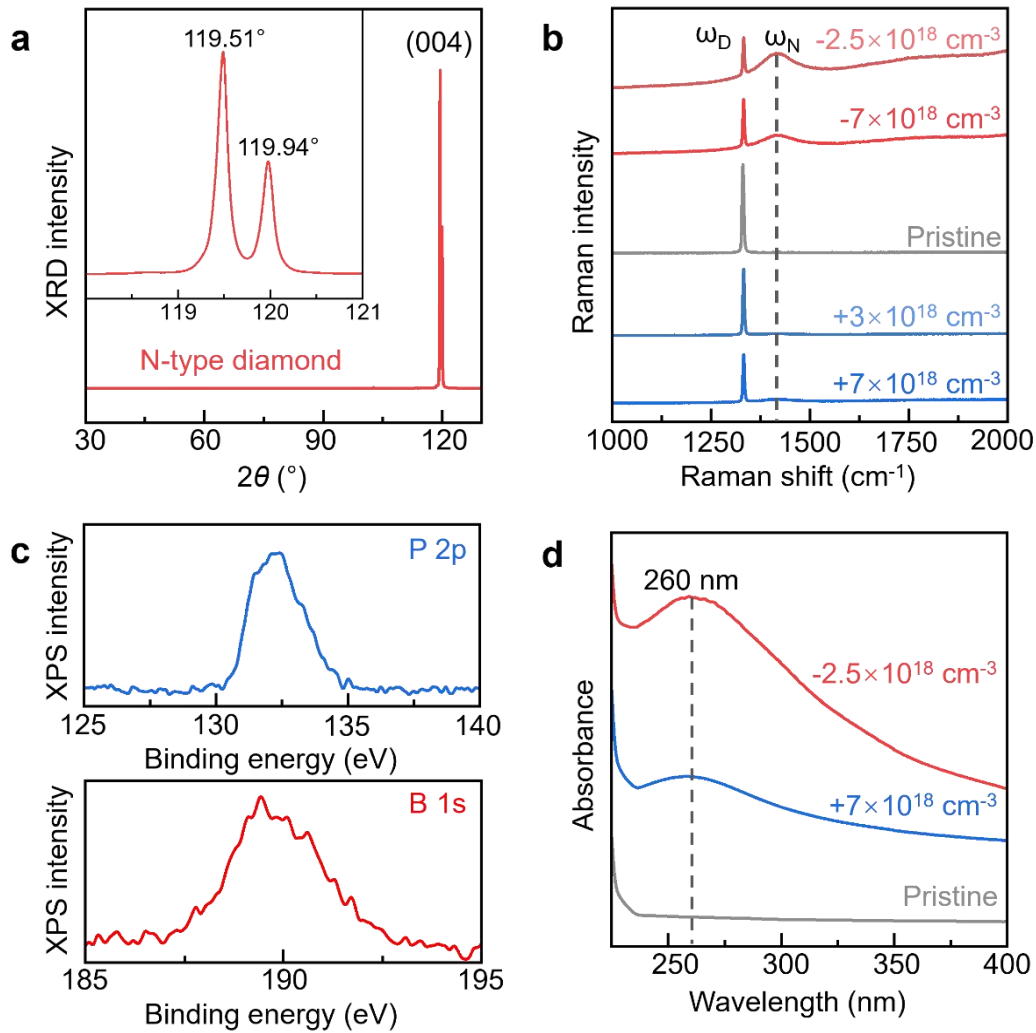

**Figure 1.** Structural and spectroscopic signatures of diamond samples. a, X-ray diffraction (XRD) pattern of the n-type diamond sample D250114, confirming the diamond phase. b, Raman spectra of representative n-type diamond, p-type diamond and pristine diamond substrate, highlighting the doping-dependent evolution of the Raman features. c, X-ray photoelectron spectroscopy (XPS) spectra of the n-type diamond sample D241125, showing

the presence of dopant-related core-level signals (P 2p and B 1s). d, Ultraviolet absorption spectra of diamond samples, revealing the codoping-induced modification of the absorption edge.

**Table 2 The $I_{\omega_N}/I_{\omega_D}$ for the diamond samples.**

| Sample | D241124 | D251010 | D250924 | D251009 |
|---|---|---|---|---|
| Doping type | N | N | P | P |
| Carrier concentration (cm$^{-3}$) | - 2.53×10$^{18}$ | - 7.14×10$^{18}$ | +3.57×10$^{18}$ | +7.19×10$^{18}$ |
| $I_{\omega_N}/I_{\omega_D}$ | 10.83 | 3.76 | 0.246 | 0.767 |

Figure. 2a and 2b show the SIMS results of the n-type and p-type diamond samples doped with both boron and phosphorus. In Figure. 2a, the n-type diamond sample D250114 exhibits a boron concentration of $1.34×10^{19}$ cm$^{-3}$, a phosphorus concentration of $3.46×10^{17}$ cm$^{-3}$, and a hydrogen concentration of $4.01×10^{19}$ cm$^{-3}$, with all impurity concentrations remaining uniform within a depth range of 0-2 μm. In contrast, Figure. 2b shows that the p-type diamond sample D250924 has a boron concentration of $2.34×10^{19}$ cm$^{-3}$, a phosphorus concentration of $7.80×10^{17}$ cm$^{-3}$, while the hydrogen concentration is below the detection limit. In the n-type diamond, the presence of hydrogen atoms and boron leads to co-doping effect, thereby forming a certain shallow-level donor. As the phosphorus doping concentration is two magnitudes lower than the Hall-effect carrier concentration, the n-type behavior should not be attributed to phosphorus doping alone; thus, the complex formed by boron and hydrogen should be the donor responsible for the n-type behavior. It has been noted that the BH$_2$ complex can theoretically serve as a shallow donor center in diamond[20]. It has also been reported that deuterium treatment of boron-doped diamond can transform the original p-type conductivity into n-type conductivity[22]. Although this deuterium-induced n-type conductivity remains under debate, many subsequent experiments have confirmed the fundamental possibility of this transformation[28]. In our case, the $BH_2$ complex is the only possibility for the formation of the n-type diamond as only the concentration of hydrogen is over $4.01 \times 10^{19}$ cm$^{-3}$ larger than the Hall effect electron concentration of the D250114 sample.

Figure. 2d presents the cathodoluminescence (CL) spectra of the n-type diamond sample D241124, the p-type diamond sample D251009 and the substrate. Since diamond is an indirect bandgap semiconductor, its intrinsic band-edge emission located at approximately 225 nm cannot be effectively detected by CL measurements; however, the luminescence is highly enhanced by doping both for the n-type diamond and p-type diamond. As shown in Figure. 2d, a pronounced emission peak is clearly observed at 273.8 nm, corresponding to an energy of about 4.51 eV, which can be attributed to the mixed peak of exciton related recombination centers, as a result of band gap narrowing, consistent with the absorption measurements and theoretical prediction[17].The strong band-edge CL emission around 280nm is repeatable confirmed and the emission intensity is not correlated with the conductivity type as shown in Figure. S3, which means that the band-edge emission is highly correlated with the impurity formation predicted by the theory[17]. .In the first step, H atoms passivate B acceptors in a 1:1 ratio, forming passive (B+H) complexes. Each B atom consumes exactly one H atom, thus for a B concentration of [B], at least that same amount of H is needed to complete passivation and create the impurity bands. In the second step, the H concentration must exceed the B concentration and the excess hydogen is available to act as effective donors that dope the impurity bands. These excess H atoms bind to existing (B+H) complexes, forming (H-B-H) triplets that produce the shallow donor levels. So the essential relationship is [H]>[B], and the effective donor concentration is approximately [H]−[B].This has an interesting dual role for B concentration. On one hand, higher [B] means more (B+H) complexes, which broadens the impurity band and reduces ionization energy. On the other hand, higher [B] also means you need more total [H] before any excess becomes available for doping. High concentration of boron doping promotes the formation of a well-developed impurity band above the percolation threshold, however more hydrogen atoms are needed to surpass that concentration of boron and provide a meaningful donor  doping. As shown in Figure. 2e and Figure. 2f, the comparsion between the energy diagram of the p-type diamond and n-type diamond is obvious, where the donor level is contributed by the excess hydrogen, consistent with the SIMS results shown in Figure. 2a and Figure. 2b and also the theoretical prediction[17].

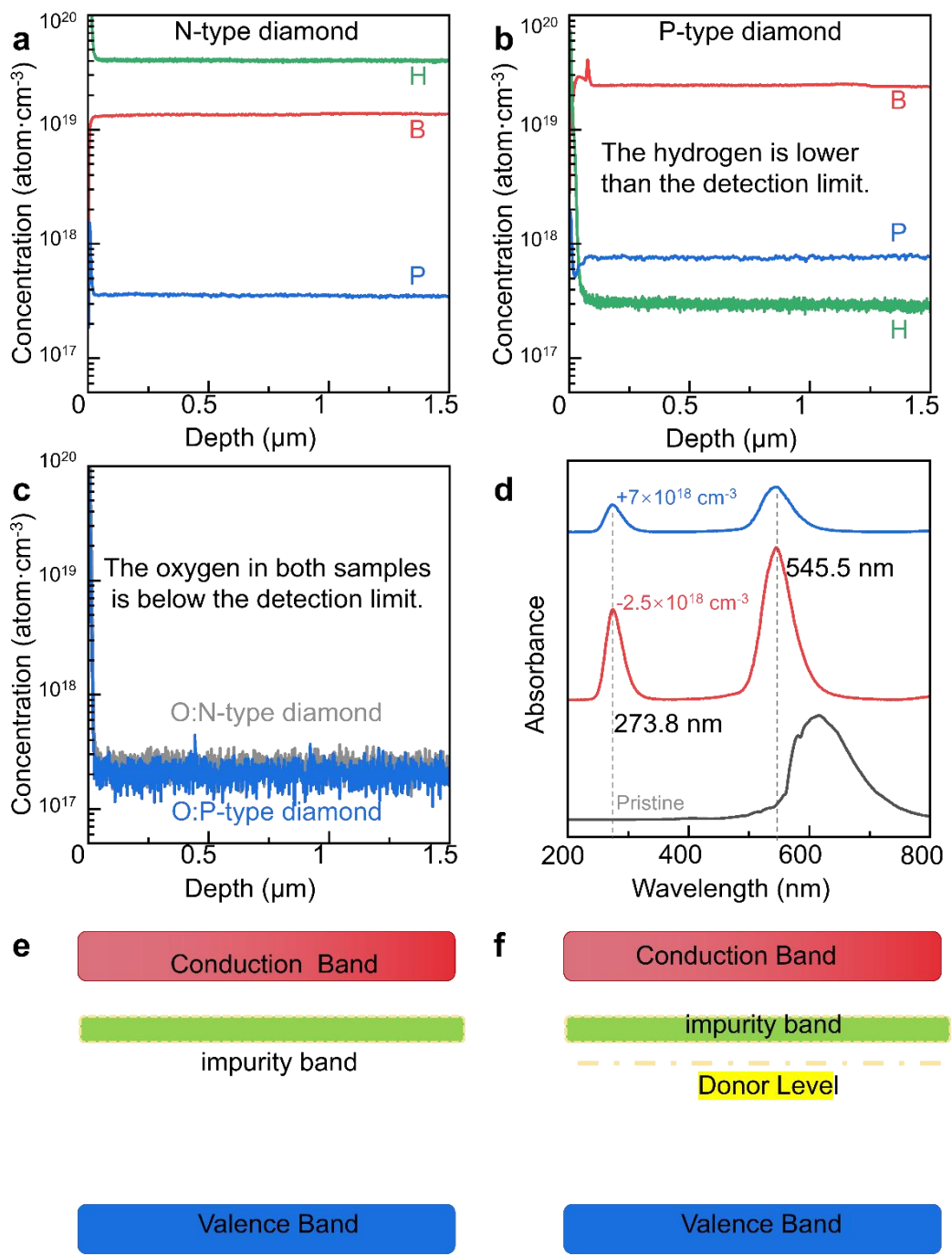

**Figure 2.** Dopant depth profiles and cathodoluminescence signatures of diamond samples. a, Secondary-ion mass spectrometry (SIMS) depth profiles of B, P and H in the n-type diamond D250114. b, SIMS depth profiles of B, P and H in the p-type diamond D250924. c, SIMS oxygen profiles for the two diamond samples, with oxygen signals remaining below the detection limit under the measurement conditions. d, CL spectra of n-type diamond D241124,

p-type diamond D251009 and pristine diamond substrate. d, energy diagram of p-type diamond D251009. e, energy diagram of n-type diamond D241124 with donor level contributed by excess of hydrogen.

Figure. 3a presents the temperature-dependent ultraviolet (UV) photoluminescence (PL) spectra of the co-doped n-type diamond. Three reproducible emission features are observed at 269.8 nm (peak A), 277.6 nm (peak B), and 285.3 nm (peak C). Their spectral positions remain essentially unchanged over the measured temperature range, indicating that the underlying radiative transitions are relatively insensitive to thermal lattice expansion and are instead governed by well-defined electronic states introduced by the co-doping chemistry. To benchmark these emissions against control samples, Figure. 3b compares the room-temperature PL of the n-type diamond with those of p-type controls (with and without phosphorus) and a pristine diamond substrate. The co-doped n-type diamond exhibits markedly enhanced UV emission, whereas the pristine substrate shows negligible PL and the p-type controls are substantially weaker. This comparison highlights that the strong UV luminescence is specifically associated with the co-doped n-type conFigureuration rather than being a generic feature of the substrate or boron-only incorporation. Notably, the boron-doped sample without phosphorus incorporation displays very weak UV emission compared with both the n-type and p-type samples, suggesting that phosphorus doping also plays a critical role in the formation of the radiative centers. We further evaluate the room-temperature internal quantum efficiency (IQE) using the low-temperature emission at 10 K as a radiative-limit reference (Figure. 3c). By integrating the PL intensity and taking the ratio of the 300 K to 10 K values, the IQE is estimated to reach 69.4%, confirming that radiative recombination remains efficient even at room temperature in this co-doped diamond system. To quantify thermal quenching, Figure. 3d summarizes the temperature-dependent integrated intensities of the three emission peaks and fits them with a thermal-activation model. The extracted activation energies are 125.7 meV (285.3 nm), 61.6 meV (277.6 nm), and 82 meV (269.8 nm), indicating distinct nonradiative escape or ionization pathways for the three radiative channels.

Notably, the 82 meV activation energy of peak A provides an experimentally accessible energy scale consistent with the free-exciton binding energy of diamond (~80 meV), allowing us to ascribe the 269.8 nm emission to a narrow-gap free-excitonic transition at 4.60 eV. The effective bandgap can accordingly be estimated as 4.60 + 0.082 = 4.682 eV, which is close to the previously determined diamond bandgap of ~4.77 eV obtained from absorption spectroscopy. Given the abundance of H-B-H donor complexes revealed by our structural characterization, we ascribe the 277.6 nm (4.47 eV) emission to a donor-bound-to-valence-band transition ($D^0h$). The spectral donor depth, defined as the energy difference between the free-excitonic transition energy and the emitted photon energy, yields a donor level of $E_d = 4.60 - 4.47 = 0.13$ eV. This value is notably twice the thermal activation energy of 61.6 meV. The relationship $n \propto \exp(-E_d/2kT)$ implies that the Fermi level is pinned approximately midway between the donor state and the impurity-band minimum (IBM) edge, consistent with a high electron concentration at room temperature as confirmed by Hall-effect measurements. The coordination of boron and hydrogen creates a mutually passivated impurity band (MPIB), effectively lowering the transport threshold to an IBM and yielding a narrowed effective bandgap of ~4.7 eV. Within this framework, the (H-B-H) triplet complex acts as a shallow donor by providing electrons to the MPIB. The shallow nature of this donor state is governed by a critical structural relaxation in which a hydrogen atom migrates from an antibonding (AB) site to a bond-center (BC) site. Phosphorus is proposed to serve as a local strain modulator: the larger phosphorus atom creates a local lattice environment that makes it energetically more favorable for the hydrogen atom in a neighboring H-B-H complex to undergo the AB-to-BC structural relaxation.

The 285.3 nm (4.35 eV) emission is attributed to a free-electron-to-acceptor (eA$^0$) transition. Following the same analytical approach used for the donor level, the spectral acceptor depth is calculated as $E_A = 4.60 - 4.35 = 0.25$ eV, which is in accordance with that of boron related acceptors. Remarkably, the thermal activation energy of 125.7 meV extracted from Figure. 3d is approximately half of this spectral acceptor depth ($E_{A/2}$ =125 meV), in excellent agreement with the relationship $p \propto \exp(-E_A/2kT)$. This self-consistent factor-of-two correspondence between the spectral depth and the thermal activation energy—observed for both the donor ($E_d$=0.13 eV, $E_{act} = 61.6$ meV) and the acceptor ($E_A$=0.25 eV, $E_{act} = 125.7$ meV)—provides strong mutual validation of the energy-level assignments and further corroborates the MPIB framework. The acceptor level at 0.25 eV above the valence band is attributed to the boron acceptor state, whose depth is modified from the conventional ~0.37 eV value in undoped diamond due to the band-gap renormalization induced by the MPIB. Figure. 3e illustrates the energy-band structure of the co-doped diamond, in which the peaks at 4.60 eV, 4.47 eV, and 4.35 eV have been unambiguously assigned.

It is also evident from Figure. 3d that the intensities of the 4.47 eV (D$^0$h) and 4.35 eV (eA$^0$) peaks first increase and then decrease with rising temperature up to room temperature, exhibiting a non-monotonic trend. The initial rise in intensity confirms the presence of an impurity band in which carriers are relatively immobile at ~50 K but become increasingly active as they are thermally excited into more delocalized states within the impurity band by overcoming potential barriers. The subsequent decrease at higher temperatures results from thermal ionization of the shallow levels as kT approaches the respective activation energies of 61.6 meV and 125.7 meV. Returning to Figure. 3c, the high IQE of 69.4% can be understood as a direct consequence of the MPIB. The MPIB not only provides a high density of radiative states but also effectively passivates nonradiative deep-level defects. This structural optimization allows the (H-B-H) triplet complexes to maintain high radiative recombination rates even at room temperature, as evidenced by the minimal thermal quenching observed between 10 K and 300 K (Figure. 3a). At the hydrogen concentration employed here ($4.01 \times 10^{19}$ cm$^{-3}$), the n-type conductivity is effectively supported by trace phosphorus doping ($3.46 \times 10^{17}$ cm$^{-3}$ ). We propose that phosphorus stabilizes the H-B-H triplet complexes by modulating local lattice strain, thereby facilitating the crucial AB-to-BC structural relaxation and sustaining the high IQE of 69.4% at room temperature.In summary, the temperature-dependent PL measurements not only reveal bandgap renormalization but also uniquely disclose the existence of shallow donors in this novel diamond system[17].

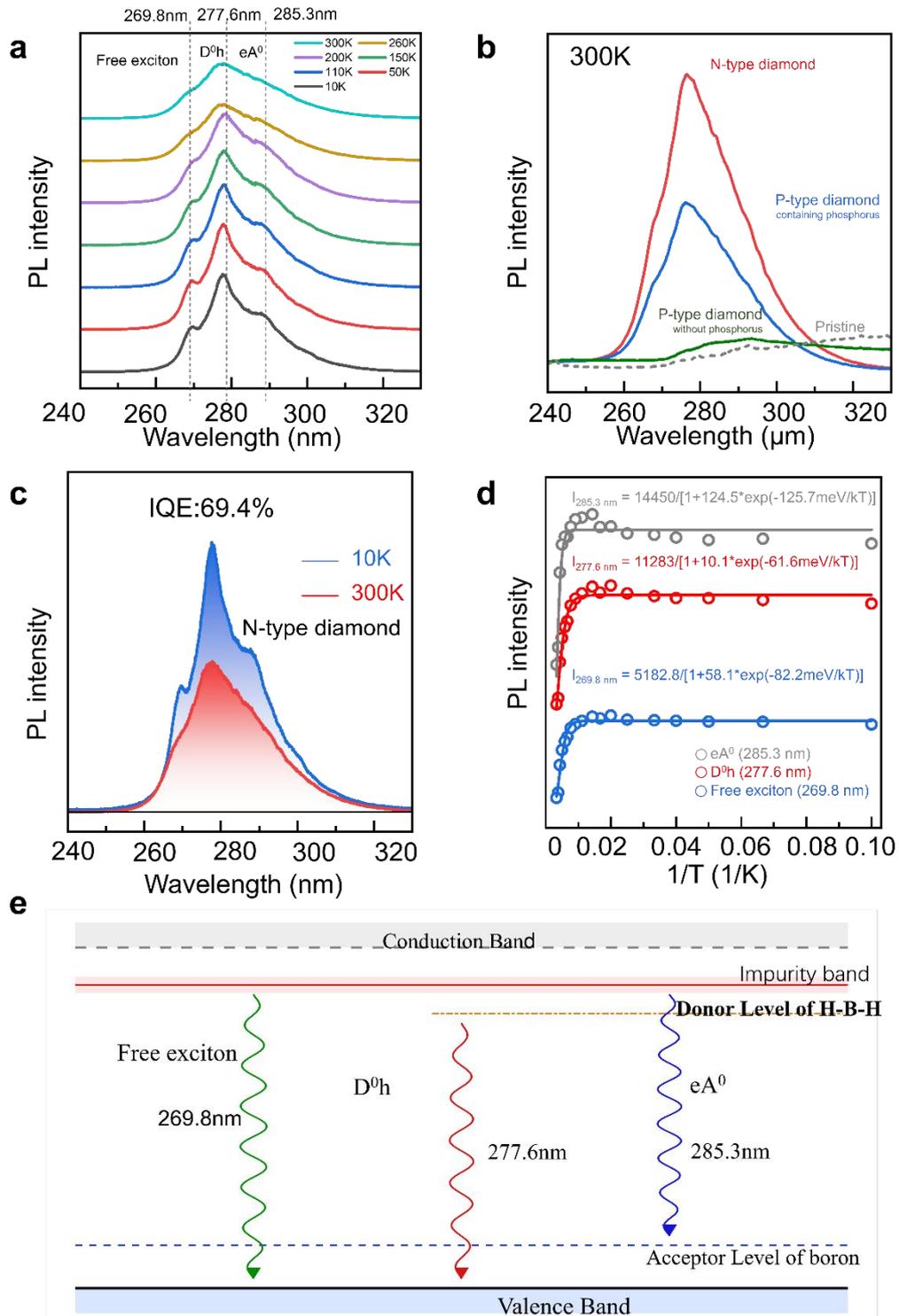

**Figure 3.** Photoluminescence characteristics and proposed optical transitions of the n-type

diamond. a, Temperature-dependent ultraviolet PL spectra of the n-type diamond, revealing three reproducible emission features at 269.8 nm, 277.6 nm, and 285.3 nm (indicated by dashed lines). b, PL comparison among the n-type diamond, p-type diamonds (boron-doped, with and without phosphorus co-doping), and pristine diamond at 300 K. c, PL spectra of the n-type diamond acquired at 10 K and 300 K. By referencing the low-temperature emission as the radiative-limit baseline, the internal quantum efficiency (IQE) is estimated to be 69.4% from the integrated PL intensity ratio. d, Temperature-dependent integrated PL intensities of the emissions at 269.8 nm (free exciton), 277.6 nm (eA$^0$), and 285.3 nm (D$^0$h), with thermal-quenching fits. e, Schematic band diagram illustrating the band-gap-narrowed excitonic emission (4.60 eV, 269.8 nm) and impurity- and complex-related transitions (4.47 eV, 277.6 nm; 4.35 eV, 285.3 nm) compared with the intrinsic diamond band gap (5.47 eV).

As shown in Figure. 4, the microstructural characterization confirms that the n-type diamond sample D250114 maintains high crystalline quality. The high-resolution TEM (HRTEM) bright-field image (Figure. 4a1) exhibits well-resolved and continuous lattice fringes with measured interplanar spacings of ~0.184 nm and ~0.212 nm, corresponding to the {200} and {111} planes of cubic diamond, respectively. Both values are slightly larger than the theoretical d-spacings of 0.178 nm and 0.206 nm, suggesting a modest lattice expansion possibly induced by phosphorus incorporation. As shown in the Figure. S4, the lattice expansion can also found in p-type diamond, confirming the phosphorus doping can induce increase the lattice parameter. The corresponding HRTEM dark-field image (Figure. 4a2) reveals consistent lattice fringes with comparable d-spacings (~0.185 nm and ~0.212 nm), further confirming a well-crystallized and uniformly oriented domain with relatively homogeneous contrast. The selected-area electron diffraction (SAED) pattern acquired along the [110] zone axis (Figure. 4b) displays sharp and discrete diffraction spots with indexed (111) and (200) reflections, confirming the single-crystalline nature of the diamond lattice without discernible signatures of secondary phases or polycrystalline rings.

To probe dopant incorporation, EELS measurements were performed at three representative positions across the TEM lamella (Figure. 4c). The stacked spectra (Figure. 4d$_1$-d$_2$) exhibit consistent C K-edge features at all positions, indicating a uniform carbon matrix within the analyzed area. Importantly, the spectra also reveal discernible boron K-edge (~200 eV) and phosphorus L-edge (~132 eV) signals, demonstrating the co-incorporation of boron and phosphorus in the n-type diamond[29,30]. More detailed EELS spectrum is shown in  Figure. S5, where the boron K-edge signal is much higher for the p-type sample, consistent with the SIMS and Hall-effect measurements. The detection of both dopant-related edges at multiple spatial positions, combined with the reproducibility of the spectral line shapes among different acquisition points, supports a spatially uniform distribution of boron and phosphorus within the probed lamella, rather than isolated local contamination or segregated precipitates. Collectively, these results confirm that the n-type diamond D20250114 preserves excellent crystallinity while incorporating B- and P-related dopant species with an overall uniform distribution at the microscale.

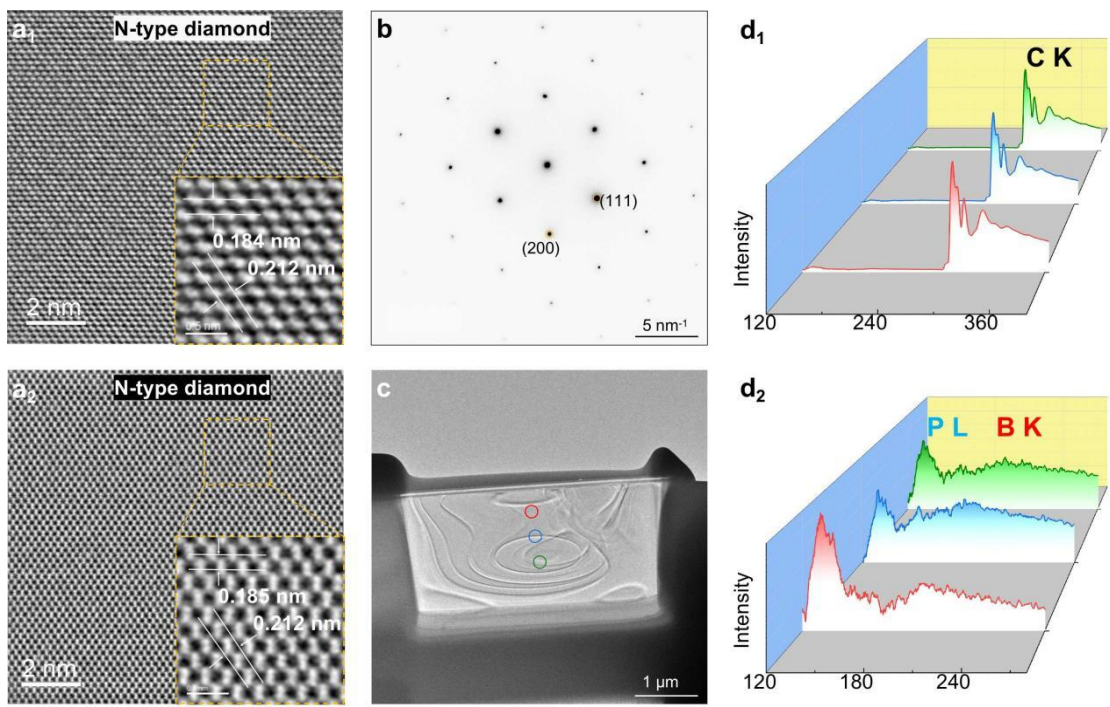

**Figure 4.** Microstructure and elemental evidence of n-type diamond. a, Bright-field ($a_1$) and dark-field ($a_2$) HRTEM images of the n-type diamond (sample D250114), exhibiting high crystallinity. Insets show magnified views with measured lattice fringe spacings of ~0.184 nm and ~0.212 nm, corresponding to the {200} and {111} planes of cubic diamond, respectively. b, Selected-area electron diffraction (SAED) pattern along the [110] zone axis, with (111) and (200) reflections indexed. c, Low-magnification TEM image of the FIB-prepared lamella; colored circles (red, blue, green) mark the EELS acquisition positions. d, EELS spectra acquired at the positions indicated in c: C K-edge ($d_1$) and B K-edge and P L-edge ($d_2$), confirming the consistent presence of both boron and phosphorus across all sampled regions.

**3. Conclusion**

In summary, we have realized stable n-type conductivity in boron-hydrogen-phosphorus co-doped single-crystalline diamond grown by MPCVD, achieving electron concentrations exceeding $1 \times 10^{19}$ cm$^{-3}$ with resistivities as low as 0.249 $\Omega \cdot$cm. Comprehensive Hall-effect measurements, together with Raman, XRD, XPS, SIMS, CL, and TEM characterization, have confirmed the tri-elements doped single crystalline n-type diamond has been achieved. The phosphorus concentration is significantly lower than the measured electron concentration, while hydrogen is present at high levels only in the n-type samples, strongly indicating that BH$_x$ complexes, rather than isolated phosphorus donors, are responsible for the observed shallow donor behavior. The phosphorus doping should facilitate the formation of BH$_x$ complexes. This tri-element co-doping route offers a promising pathway for realizing diamond-based p-n junctions and diamond chips.

## 4. Experimental Section

**Diamond synthesis:** Single crystal diamond is grown by MPCVD using $CH_4$ and $H_2$ as the main source gases, and doped sources of borane ($BH_3$) and phosphine were added to form a mixed gas for growth under the input power of 5 kW, where the pressure of the cavity and surface temperature of the seed crystal were maintained at 100 Torr and 930 °C. After growth, the single-crystal doped diamond was sequentially sonicated with acetone, isopropanol, ethanol and deionized water for 5 minutes to remove residues from the surface. The samples were then blow-dried with a $N_2$ air gun.

**Raman, XRD, Hall test, SIMS, and UV absorption spectroscopy:** The laser Raman spectrometer (inVia-Reflex) was used for Raman characterization of the diamond samples, using a 532 nm laser source to scan the spectrum from 500 $cm^{-1}$ to 2000 $cm^{-1}$. Diamond samples were measured using XRD (Bruker D8 ADVANCE) and scanned at 0.02° in the range of 0° to 130°. The Hall test instrument (Nanometrics HL5500) was used to determine the carrier concentration, resistivity and Hall mobility of single crystal diamond. The SIMS test instrument (CAMECA IMS 7f-Auto) was used to measure the distribution of boron, phosphorus, hydrogen, and oxygen at different depths of diamond. By adjusting the spot radius of the ion beam, the concentration of bulk diamond impurities can be effectively characterized. The ultraviolet absorption spectra of the diamond samples were measured using a UV-visible spectrophotometer (Lambda 950).

**PL and CL spectroscopy:** PL spectra of the diamond samples were measured using Horiba iHR550 spectrometer equipped with a CCD detector. The 405 nm laser was used as the excitation source, and the beam size of focused laser approached 1 μm with a 100× objective. The CL emission spectra were registered at room temperature using a grating spectrometer equipped with a charge-coupled device camera. It was recorded by the grating StellarNet's

SILVVER-Nova (StellarNet, Inc. Tampa, Fl, USA) high performance fiber optic spectrometer covering the 190-1100 nm wavelength range equipped with a TE cooled 2048 pixel charge coupled device (CCD) camera. Two gratings were used, giving a spectral resolution of 0.5 nm or 0.2 nm in the case of high-resolution spectra. The accelerating voltage of the electron beam for the CL spectrum excitation was set at 30 kV.

**TEM and EELS measurements:** In order to prepare TEM sample, the n-type diamond was precoated with Pt layers, and then shaped into an ultrathin slice with a thickness of 50-60 nm by focused ion beam (Thermo Fisher SCIOS 2). TEM and SAED measurements were conducted by HITACHI HF5000 with an acceleration voltage of 200 kV. Meanwhile, EELS spectra around the C K-edge, B K-edge and P K-edge were collected by Gatan 1069.

**Acknowledgements**

S.S.L. thanks the support from the National Natural Science Foundation of China (No. 51202216, 51551203, and 61774135), Youth Talent from the Central Organization Department.

**Author contributions**

S.L. designed the experiments and carried out the experiments, analyzed the data, discussed the results and wrote the paper. H.B. and S.H. carried out the experiments, analyzed the data and wrote the paper. F.W. et al. discussed the results and participated in experiments. All authors contributed to the preparation of the manuscript.

**Competing interests**

The authors declare no competing interests.

**Data Availability Statement**

The data that support the findings of this study are available from the corresponding author upon reasonable request.

**Keywords**

n-type diamond, co-doping, diamond heterojunction devices.

**Uncategorized References**

[1]    Isberg J, Hammersberg J, Johansson E, Wikström T, Twitchen D J, Whitehead A J, Coe S E and Scarsbrook G A 2002 High Carrier Mobility in Single-Crystal Plasma-Deposited Diamond *Science (1979).* **297** 1670–2

[2]    Yan C, Vohra Y K, Mao H and Hemley R J 2002 Very high growth rate chemical vapor deposition of single-crystal diamond *Proceedings of the National Academy of Sciences* **99** 12523–5

[3]    Nesladek M 2005 Conventional n-type doping in diamond: state of the art and recent progress *Semicond. Sci. Technol.* **20** R19

[4]    Koizumi S, Watanabe K, Hasegawa M and Kanda H 2001 Ultraviolet Emission from a Diamond pn Junction *Science (1979).* **292** 1899–901

[5]    Gheeraert E, Casanova N, Tajani A, Deneuville A, Bustarret E, Garrido J A, Nebel C E and Stutzmann M 2002 n-Type doping of diamond by sulfur and phosphorus *Diam. Relat. Mater.* **11** 289–95

[6]    Lin S S 2012 Robust low resistivity p-type ZnO:Na films after ultraviolet illumination: The elimination of grain boundaries *Appl. Phys. Lett.* **101** 122109

[7]    Amano H, Kito M, Hiramatsu K and Akasaki I 1989 P-Type Conduction in Mg-Doped GaN Treated with Low-Energy Electron Beam Irradiation (LEEBI) *Jpn. J. Appl. Phys.* **28** L2112–4

[8]    Goss J P, Briddon P R, Sque S J and Jones R 2004 Boron-hydrogen complexes in diamond *Phys. Rev. B* **69** 165215

[9]    Liu D Y, Hao L C, Teng Y, Qin F, Shen Y, Tang K, Ye J D, Zhu S M, Zhang R, Zheng Y D and Gu S L 2021 Nitrogen modulation of boron doping behavior for accessible n-type diamond *APL Mater.* **9** 081106

[10] Ekimov E A, Sidorov V A, Bauer E D, Mel'nik N N, Curro N J, Thompson J D and Stishov S M 2004 Superconductivity in diamond *Nature* **428** 542–5

[11] Lin S, Yu X, Yang M, Zhong H, Guo J and Chen X 2024 Superconductivity and Metallic Behavior in Heavily Doped Bulk Single Crystal Diamond and Novel Transportation Behavior at Graphene/Diamond Heterostructure *Adv. Funct. Mater.* **34** 2410876

[12] Ishii N and Shimizu T 1998 Cluster-model calculation of hyperfine parameters for the $\langle 100 \rangle$-split $[\mathrm{B}\ensuremath{-}\mathrm{N}]^{+}$ interstitial in diamond *Phys. Rev. B* **58** 12533–5

[13] Isoya J, Kanda H, Akaishi M, Morita Y and Ohshima T 1997 ESR studies of incorporation of phosphorus into high-pressure synthetic diamond *Diam. Relat. Mater.* **6** 356–60

[14] Fan K, Tang K, Zhang M, Wu K, Zhao G, Huang Y, Zhu S, Ye J and Gu S 2023 The boron-phosphorous co-doping scheme for possible n-type diamond from first principles *Comput. Mater. Sci.* **222** 112113

[15] Lombardi E B 2009 Boron–hydrogen complexes in diamond and their vibrational properties *Diam. Relat. Mater.* **18** 835–8

[16] Lin S S, Lu J G, Ye Z Z, He H P, Gu X Q, Chen L X, Huang J Y and Zhao B H 2008 p-type behavior in Na-doped ZnO films and ZnO homojunction light-emitting diodes *Solid State Commun.* **148** 25–8

[17] Yan Y, Li J, Wei S-H and Al-Jassim M M 2007 Possible Approach to Overcome the Doping Asymmetry in Wideband Gap Semiconductors *Phys. Rev. Lett.* **98** 135506

[18] Nakamura S, Iwasa N, Masayuki Senoh M S and Takashi Mukai T M 1992 Hole Compensation Mechanism of P-Type GaN Films *Jpn. J. Appl. Phys.* **31** 1258

[19] Neugebauer J and Van de Walle C G 1996 Role of hydrogen in doping of GaN *Appl. Phys. Lett.* **68** 1829–31

[20]   Dai Y, Dai D, Liu D, Han S and Huang B 2004 Mechanism of p-type-to-n-type conductivity conversion in boron-doped diamond *Appl. Phys. Lett.* **84** 1895–7

[21]   Kumar A, Pernot J, Omnès F, Muret P, Traoré A, Magaud L, Deneuville A, Habka N, Barjon J, Jomard F, Pinault M A, Chevallier J, Mer-Calfati C, Arnault J C and Bergonzo P 2011 Boron-deuterium complexes in diamond: How inhomogeneity leads to incorrect carrier type identification *J. Appl. Phys.* **110** 033718

[22]   Teukam Z, Chevallier J, Saguy C, Kalish R, Ballutaud D, Barbé M, Jomard F, Tromson-Carli A, Cytermann C, Butler J E, Bernard M, Baron C and Deneuville A 2003 Shallow donors with high n-type electrical conductivity in homoepitaxial deuterated boron-doped diamond layers *Nat. Mater.* **2** 482–6

[23]   Das D, Kandasami A and Ramachandra Rao M S 2021 Realization of highly conducting n-type diamond by phosphorus ion implantation *Appl. Phys. Lett.* **118** 102102

[24]   Liu X, Chen X, Singh D J, Stern R A, Wu J, Petitgirard S, Bina C R and Jacobsen S D 2019 Boron–oxygen complex yields n-type surface layer in semiconducting diamond *Proceedings of the National Academy of Sciences* **116** 7703–11

[25]   Pu M, Zhang F, Liu S, Irifune T and Lei L 2019 Tensile-strain induced phonon splitting in diamond* *Chinese Physics B* **28** 053102

[26]   Cheng C, Sun X, Shen W, Wang Q, Li L, Dong F, Liang K and Wu G 2024 Enhancing n-type doping in diamond by strain engineering *J. Phys. D Appl. Phys.* **57** 485103

[27]   Saguy C, Kalish R, Chevallier J, Jomard F, Cytermann C, Philosoph B, Kociniewski T, Ballutaud D, Baron C and Deneuville A 2007 The p-to-n-type conversion of boron-doped diamond layers by deuteration: New findings *Diam. Relat. Mater.* **16** 1459–62

[28]   Dey R, Dolai S, Hussain S, Bhar R and Pal A K 2018 Phosphorus doping of diamond-like carbon films by radio frequency CVD-cum-evaporation technique *Diam. Relat. Mater.* **82** 70–8

[29]   Lu Y-G, Turner S, Verbeeck J, Janssens S D, Wagner P, Haenen K and Van Tendeloo G 2012 Direct visualization of boron dopant distribution and coordination in individual chemical vapor deposition nanocrystalline B-doped diamond grains *Appl. Phys. Lett.* **101** 041907

[30]   Wu P-H, Ku W-H, Chiu K-A and Chang L 2022 Radiation Damage in (001) Diamond Induced by Phosphorus Ion Implantation *physica status solidi (a)* **219** 2100829